\documentclass[letterpaper,aps,prl,twocolumn,showpacs,superscriptaddress]{revtex4} 

\usepackage{rotating}
\usepackage{graphicx}  
\usepackage{dcolumn}   
\usepackage{bm}        
\usepackage{amssymb}   
\usepackage{amsmath, amsthm}

\newcommand{\pp}  {\mbox{$p\bar{p}$}}
\newcommand{\roots}  {\mbox{$\sqrt{s}=1.96$~TeV}}
\newcommand{\zmumu}  {\mbox{$Z/\gamma^*\rightarrow\mu^+ \mu^-$}}
\newcommand{\zee}  {\mbox{$Z/\gamma^*\rightarrow e^+e^-$}}
\newcommand{\zmm}  {\mbox{$Z/\gamma^*\rightarrow\mu^+ \mu^-$}}

\newcommand{\ztau}  {\mbox{$Z/\gamma^*\rightarrow\tau^+ \tau^-$}}
\newcommand{\z}  {\mbox{$Z/\gamma^*$}}
\newcommand{\Z}  {\mbox{$Z/\gamma^*$}}
\newcommand{\W}  {\mbox{$W$}}
\newcommand{\w}  {\mbox{$W$}}
\newcommand{\pt}  {\mbox{$p_{T}$}}
\newcommand{\ptz}  {\mbox{$p_{T}^{Z}$}}

\newcommand{\fewz}  {{\sc fewz}}
\newcommand{\dynnlo}  {{\sc dynnlo}}
\newcommand{\resbos}  {{\sc resbos}}
\newcommand{\photos}  {{\sc photos}}
\newcommand{\herwig}  {{\sc herwig}}
\newcommand{\pythia}  {{\sc pythia}}
\newcommand{\alpgen}  {{\sc alpgen}}
\newcommand{\sherpa}  {{\sc sherpa}}
\newcommand{\mcfm}  {{\sc mcfm}}
\newcommand{\geant}  {{\sc geant}}
\newcommand{\jimmy}  {{\sc jimmy}}
\newcommand{\perugia}  {{Perugia 6}}
\newcommand{\cteq}  {{CTEQ}}
\newcommand{\mstw}  {{MSTW}}
\newcommand{\comix}  {{\sc comix}}

\begin{document}

\hspace{5.2in} \mbox{FERMILAB-PUB-10-183-E}

\title{Measurement of the normalized {\boldmath $Z/\gamma^*\rightarrow\mu^+\mu^-$} transverse momentum distribution in {\boldmath $p\bar{p}$}~collisions at {\boldmath $\sqrt{s}=1.96$}~TeV}

%
\affiliation{Universidad de Buenos Aires, Buenos Aires, Argentina}
\affiliation{LAFEX, Centro Brasileiro de Pesquisas F{\'\i}sicas, Rio de Janeiro, Brazil}
\affiliation{Universidade do Estado do Rio de Janeiro, Rio de Janeiro, Brazil}
\affiliation{Universidade Federal do ABC, Santo Andr\'e, Brazil}
\affiliation{Instituto de F\'{\i}sica Te\'orica, Universidade Estadual Paulista, S\~ao Paulo, Brazil}
\affiliation{Simon Fraser University, Vancouver, British Columbia, and York University, Toronto, Ontario, Canada}
\affiliation{University of Science and Technology of China, Hefei, People's Republic of China}
\affiliation{Universidad de los Andes, Bogot\'{a}, Colombia}
\affiliation{Charles University, Faculty of Mathematics and Physics, Center for Particle Physics, Prague, Czech Republic}
\affiliation{Czech Technical University in Prague, Prague, Czech Republic}
\affiliation{Center for Particle Physics, Institute of Physics, Academy of Sciences of the Czech Republic, Prague, Czech Republic}
\affiliation{Universidad San Francisco de Quito, Quito, Ecuador}
\affiliation{LPC, Universit\'e Blaise Pascal, CNRS/IN2P3, Clermont, France}
\affiliation{LPSC, Universit\'e Joseph Fourier Grenoble 1, CNRS/IN2P3, Institut National Polytechnique de Grenoble, Grenoble, France}
\affiliation{CPPM, Aix-Marseille Universit\'e, CNRS/IN2P3, Marseille, France}
\affiliation{LAL, Universit\'e Paris-Sud, CNRS/IN2P3, Orsay, France}
\affiliation{LPNHE, Universit\'es Paris VI and VII, CNRS/IN2P3, Paris, France}
\affiliation{CEA, Irfu, SPP, Saclay, France}
\affiliation{IPHC, Universit\'e de Strasbourg, CNRS/IN2P3, Strasbourg, France}
\affiliation{IPNL, Universit\'e Lyon 1, CNRS/IN2P3, Villeurbanne, France and Universit\'e de Lyon, Lyon, France}
\affiliation{III. Physikalisches Institut A, RWTH Aachen University, Aachen, Germany}
\affiliation{Physikalisches Institut, Universit{\"a}t Freiburg, Freiburg, Germany}
\affiliation{II. Physikalisches Institut, Georg-August-Universit{\"a}t G\"ottingen, G\"ottingen, Germany}
\affiliation{Institut f{\"u}r Physik, Universit{\"a}t Mainz, Mainz, Germany}
\affiliation{Ludwig-Maximilians-Universit{\"a}t M{\"u}nchen, M{\"u}nchen, Germany}
\affiliation{Fachbereich Physik, Bergische  Universit{\"a}t Wuppertal, Wuppertal, Germany}
\affiliation{Panjab University, Chandigarh, India}
\affiliation{Delhi University, Delhi, India}
\affiliation{Tata Institute of Fundamental Research, Mumbai, India}
\affiliation{University College Dublin, Dublin, Ireland}
\affiliation{Korea Detector Laboratory, Korea University, Seoul, Korea}
\affiliation{CINVESTAV, Mexico City, Mexico}
\affiliation{FOM-Institute NIKHEF and University of Amsterdam/NIKHEF, Amsterdam, The Netherlands}
\affiliation{Radboud University Nijmegen/NIKHEF, Nijmegen, The Netherlands}
\affiliation{Joint Institute for Nuclear Research, Dubna, Russia}
\affiliation{Institute for Theoretical and Experimental Physics, Moscow, Russia}
\affiliation{Moscow State University, Moscow, Russia}
\affiliation{Institute for High Energy Physics, Protvino, Russia}
\affiliation{Petersburg Nuclear Physics Institute, St. Petersburg, Russia}
\affiliation{Stockholm University, Stockholm and Uppsala University, Uppsala, Sweden }
\affiliation{Lancaster University, Lancaster LA1 4YB, United Kingdom}
\affiliation{Imperial College London, London SW7 2AZ, United Kingdom}
\affiliation{The University of Manchester, Manchester M13 9PL, United Kingdom}
\affiliation{University of Arizona, Tucson, Arizona 85721, USA}
\affiliation{University of California Riverside, Riverside, California 92521, USA}
\affiliation{Florida State University, Tallahassee, Florida 32306, USA}
\affiliation{Fermi National Accelerator Laboratory, Batavia, Illinois 60510, USA}
\affiliation{University of Illinois at Chicago, Chicago, Illinois 60607, USA}
\affiliation{Northern Illinois University, DeKalb, Illinois 60115, USA}
\affiliation{Northwestern University, Evanston, Illinois 60208, USA}
\affiliation{Indiana University, Bloomington, Indiana 47405, USA}
\affiliation{Purdue University Calumet, Hammond, Indiana 46323, USA}
\affiliation{University of Notre Dame, Notre Dame, Indiana 46556, USA}
\affiliation{Iowa State University, Ames, Iowa 50011, USA}
\affiliation{University of Kansas, Lawrence, Kansas 66045, USA}
\affiliation{Kansas State University, Manhattan, Kansas 66506, USA}
\affiliation{Louisiana Tech University, Ruston, Louisiana 71272, USA}
\affiliation{University of Maryland, College Park, Maryland 20742, USA}
\affiliation{Boston University, Boston, Massachusetts 02215, USA}
\affiliation{Northeastern University, Boston, Massachusetts 02115, USA}
\affiliation{University of Michigan, Ann Arbor, Michigan 48109, USA}
\affiliation{Michigan State University, East Lansing, Michigan 48824, USA}
\affiliation{University of Mississippi, University, Mississippi 38677, USA}
\affiliation{University of Nebraska, Lincoln, Nebraska 68588, USA}
\affiliation{Rutgers University, Piscataway, New Jersey 08855, USA}
\affiliation{Princeton University, Princeton, New Jersey 08544, USA}
\affiliation{State University of New York, Buffalo, New York 14260, USA}
\affiliation{Columbia University, New York, New York 10027, USA}
\affiliation{University of Rochester, Rochester, New York 14627, USA}
\affiliation{State University of New York, Stony Brook, New York 11794, USA}
\affiliation{Brookhaven National Laboratory, Upton, New York 11973, USA}
\affiliation{Langston University, Langston, Oklahoma 73050, USA}
\affiliation{University of Oklahoma, Norman, Oklahoma 73019, USA}
\affiliation{Oklahoma State University, Stillwater, Oklahoma 74078, USA}
\affiliation{Brown University, Providence, Rhode Island 02912, USA}
\affiliation{University of Texas, Arlington, Texas 76019, USA}
\affiliation{Southern Methodist University, Dallas, Texas 75275, USA}
\affiliation{Rice University, Houston, Texas 77005, USA}
\affiliation{University of Virginia, Charlottesville, Virginia 22901, USA}
\affiliation{University of Washington, Seattle, Washington 98195, USA}
\author{V.M.~Abazov} \affiliation{Joint Institute for Nuclear Research, Dubna, Russia}
\author{B.~Abbott} \affiliation{University of Oklahoma, Norman, Oklahoma 73019, USA}
\author{M.~Abolins} \affiliation{Michigan State University, East Lansing, Michigan 48824, USA}
\author{B.S.~Acharya} \affiliation{Tata Institute of Fundamental Research, Mumbai, India}
\author{M.~Adams} \affiliation{University of Illinois at Chicago, Chicago, Illinois 60607, USA}
\author{T.~Adams} \affiliation{Florida State University, Tallahassee, Florida 32306, USA}
\author{G.D.~Alexeev} \affiliation{Joint Institute for Nuclear Research, Dubna, Russia}
\author{G.~Alkhazov} \affiliation{Petersburg Nuclear Physics Institute, St. Petersburg, Russia}
\author{A.~Alton$^{a}$} \affiliation{University of Michigan, Ann Arbor, Michigan 48109, USA}
\author{G.~Alverson} \affiliation{Northeastern University, Boston, Massachusetts 02115, USA}
% ...
%
%
\collaboration{The D0 Collaboration\footnote{with visitors from
$^{a}$Augustana College, Sioux Falls, SD, USA,
$^{b}$The University of Liverpool, Liverpool, UK,
$^{c}$SLAC, Menlo Park, CA, USA,
$^{d}$ICREA/IFAE, Barcelona, Spain,
$^{e}$Centro de Investigacion en Computacion - IPN, Mexico City, Mexico,
$^{f}$ECFM, Universidad Autonoma de Sinaloa, Culiac\'an, Mexico,
and 
$^{g}$Universit{\"a}t Bern, Bern, Switzerland.%
}} \noaffiliation
\vskip 0.25cm
 
\date{June 3rd, 2010}

\begin{abstract}
We present a new measurement of the \z\ transverse momentum distribution in the range 0 -- 330~GeV, in proton-antiproton collisions at \roots.
The measurement uses 0.97~fb$^{-1}$\ of integrated luminosity recorded by the D0 experiment and is the first using the \zmm$+X$\ channel at this center-of-mass energy.
This is also the first measurement of the \z\ transverse momentum distribution that presents the result at the level of particles entering the detector, minimizing dependence on theoretical models.
As any momentum of the \z\ in the plane transverse to the incoming beams must be balanced by some recoiling system, primarily the result of QCD radiation in the initial state, this variable is an excellent probe of the underlying process.
Tests of the predictions of QCD calculations and current event generators show they have varied success in describing the data.
Using this measurement as an input to theoretical predictions will allow for a better description of hadron collider data and hence it will increase experimental sensitivity to rare signals.
\end{abstract}


\maketitle

\clearpage

In the complex environment of a hadron collider, such as the Fermilab Tevatron Collider or the CERN Large Hadron Collider, the \zee\ and \zmumu\ processes are experimentally simple to identify and have little background.
Further, reconstruction of the \z\ kinematics provides an unambiguous, colorless probe of the underlying collision process.
Momentum conservation requires that any momentum component of the \z\ in the direction transverse to the incoming hadron beams (\pt) must be balanced by a recoiling system ($X$), typically the result of QCD radiation in the initial state.
The \z\ \pt\ is therefore sensitive to the nature of this radiation across a wide momentum range, making it a compelling variable and an excellent testing ground for theoretical predictions.

Several tools have been developed which give a prediction of the \z\ \pt\ distribution, from fixed order perturbative QCD (pQCD) calculations valid at high \pt, such as \mcfm~\cite{mcfm}, \fewz~\cite{fewz} and \dynnlo~\cite{dynnlo}, 
to predictions based on gluon resummation valid at low \pt~\cite{resbos1}, such as {\sc resbos}~\cite{resbos2}.
Various complete event generators are also available, including \pythia~\cite{pythia_64}, \herwig~\cite{herwig}, \alpgen~\cite{alpgen}, and \sherpa~\cite{sherpa}, 
which cover both high and low \pt\ regions by interfacing tree-level matrix element calculations to a parton shower resummation model.
Comparisons between these generators show that they differ significantly in the predicted kinematics of boson and boson+jet production, and that these predictions have a strong dependence on various adjustable internal generator parameters~\cite{mc_paper}.
Measurements of the \z\ \pt\ and other kinematic quantities in \z\ production are therefore an essential input to improve these models, which are also used to predict the properties of rare signals like the Higgs boson and its main backgrounds: \w+jets, \z+jets and diboson production.
Such improvements will result in increased experimental sensitivity to these rare signals.

Previous measurements at the Tevatron have studied the \z\ \pt\ and rapidity distributions both inclusively~\cite{cdf_zpt, d0_runi_zpt, d0_zpt, d0_zrap} and in events with at least one jet~\cite{my_zjet}.
Other measurements have focused on the kinematics of the jets in \z\ or \W\ boson events~\cite{d0_zjet, my_zjet, cdf_zjet, cdf_wjet, henrik_zjet}, of the angular correlations between the \z\ and leading jet~\cite{zjet_angles}, and of the production of \Z\ or \W\ boson in association with heavy flavor quarks~\cite{d0_zb, cdf_zb, d0_wc, cdf_wc}.
In this Letter, we describe a new measurement of the normalized inclusive \z$\rightarrow\mu^+\mu^-$\ \pt\ distribution, the first such measurement using the dimuon channel in the Tevatron run beginning 2001 (``Run II'').
The differential dimuon+$X$\ cross section is measured as a function of the dimuon \pt\ (\ptz), then normalized to the measured inclusive dimuon cross section, canceling many systematic uncertainties.
The shape of the \z\ \pt\ distribution has previously been measured in Run~II with the \zee\ channel~\cite{d0_zpt}, using a comparable integrated luminosity. 
In that result, a resummation prediction was found to be consistent with the data in the $\pt<30$~GeV region, but pQCD predictions were found to be 25\% below the data in the region $\pt>30$~GeV. 
Compared to that result, the muon channel uses a statistically independent dataset, has a different detector acceptance, and different sources of systematic uncertainty; it therefore adds important information on the \z\ \pt\ distribution and any disagreements between the data and theory predictions. 
Due to the different response of the detector to electrons and muons, there is also different sensitivity to QED final state radiation (FSR) in the dielectron and dimuon systems~\cite{les_houches}.

An important development in this analysis is the definition of the final observable: for the first time in a measurement of the \z\ \pt,  the results are presented at the level of particles entering the detector.
Previous measurements have applied theoretical factors to go from these particles to the (non-observable) \z\ by correcting for any undetected FSR, and from the measured lepton acceptance to full 4$\pi$\ coverage, correcting for undetected leptons.
These factors rely upon models of FSR and the correlation between boson rapidity and \pt.
Here, we avoid such factors and present the data in terms of an observable: the \pt\ of the dimuon system, for muons within the detector acceptance.
This approach minimizes dependence on theoretical models, and the result can be used as an unbiased test of such models.
This is also the same definition of the dimuon final state as previous D0 measurements of \z($\rightarrow\mu^+\mu^-$)+jet+$X$\ production~\cite{my_zjet, zjet_angles}, and the relationship between \ptz\ and the production of jets in the final state makes this measurement complementary to those results.

The analysis uses a dataset of \pp\ collisions at \roots,  corresponding to an integrated luminosity of $0.97\pm0.06$~fb$^{-1}$~\cite{d0lumi}\ recorded by the D0 detector between April 2002 and February 2006.
A full description of the D0 detector is available elsewhere~\cite{d0det}, and only a brief description of the components most relevant for this analysis is given here.
The \pp\ interaction region is surrounded by two tracking detectors: a silicon microstrip tracker and a scintillating fiber tracker, both housed inside a solenoidal magnet providing a field of approximately 2~T.
These trackers provide a momentum measurement for charged particles and are used to reconstruct the primary interaction point in each collision.
Outside the solenoid is a liquid-argon and uranium calorimeter which is split into three sections: a central section covering $|\eta|<1.1$~\cite{eta} and two forward sections covering $1.4<|\eta|<4.2$.
Outside the calorimeter there are three layers of muon detectors, made of a combination of scintillation counters and drift chambers covering $|\eta|<1$, and scintillation counters and drift tubes extending the coverage to $|\eta|<2$.
A 1.8~T iron toroidal magnet is located between the first and second layer, providing an independent momentum measurement for muons.

Events used in this analysis are selected by at least one of a suite of single-muon triggers. 
These triggers used a fast readout from muon system, or a combination of the fiber tracker and muon system, to identify muon candidates.
Then information from the full tracking and muon systems is incorporated to provide further rejection.
Additional requirements are then applied to the events selected by the trigger to obtain a sample of \z\ candidates.
Using the full information from the muon detectors and the tracking system, two muons of opposite charge and \pt$>15$~GeV are required, with a dimuon mass in the range $65<M_{\mu\mu}<115$~GeV.
To reject cosmic rays and poorly reconstructed muons, the muon tracks are required to be consistent with the reconstructed primary interaction point both along the beam direction and in the transverse plane.
The two muon tracks are also required not to be colinear, and to be consistent with the \pp\ bunch crossing time using timing information from the muon system scintillators.
Further selections are applied to limit the measurement to regions with high detection efficiency: the muons are required to have $|\eta|<1.7$, and the primary vertex must lie within 50~cm of the center of the detector in the coordinate along the direction of the beam.
In total, 59,336 dimuon candidate events pass all selection requirements.

The main background in this analysis is  dijet production with two semi-leptonic decays, or $W$+jets production in which one muon comes from the $W$\ and the other from a semi-leptonic decay in a jet.
These events are reduced to a negligible level by two isolation requirements. 
First, we reject overlaps between muons and jets with \pt$>15$~GeV, by requiring angular separation $\sqrt{(\Delta\phi)^2 + (\Delta\eta)^2} > 0.5$, where $\phi$\ is the azimuthal angle.
Then, we require the product of the isolation variables for the two  muons to be $<0.05$, where each isolation variable is calculated by taking the sum of track \pt\ and calorimeter energy in a cone around each muon (excluding the muon track and calorimeter energy associated with the muon itself), and dividing by the muon \pt.

The remaining contribution from these backgrounds is estimated from data by studying the product of the isolation variables for muons failing isolation requirements.
Extrapolating into the selected region shows this background to be $<0.5$\%\ of the final sample.
The remaining backgrounds (from $t\bar{t}$, $WW$, $WZ$, and \ztau\ production), as well as the \zmm$+X$\ signal, are modeled with \pythia\ v6.409.
A separate \zmm$+X$\ sample is generated using \alpgen\ v2.11 with \pythia\ v6.409 for parton showering.
All simulated signal and background samples are normalized to higher order theoretical predictions~\cite{incl_wz, top_cs} and passed through a \geant~\cite{geant} simulation of the D0 detector.
The total background from all sources is found to be below 2\% everywhere and less than 1\% in the region \ptz$<50$~GeV. 
The estimated background contribution is subtracted from data and a 10\% systematic uncertainty is assigned to each background normalization to cover all sources of uncertainty.

To extract the shape of the \ptz\ distribution, the measured dimuon candidates must be corrected for detector resolution and efficiency, both of which are derived directly from data.
The detector resolution is extracted from the shape and position of the \z\ resonance peak in dimuon data, which is dominated by detector resolution rather than the natural width of the $Z$\ boson.
The resolution is well described by a double-Gaussian function form in $1/\pt$, with  the majority (98\%) of muons having a $1/\pt$ resolution of approximately 0.0018~GeV$^{-1}$, and the remaining 2\% (chosen at random) a  $1/\pt$ resolution of approximately 0.012~GeV$^{-1}$.

The detector efficiencies are derived using the ``tag and probe'' method on dimuon candidate pairs.
The ``tag'' muon is selected, and must pass all selection requirements, which may be adjusted as needed to remove backgrounds. 
The ``probe'' muon is then selected with one explicit reconstruction requirement removed; the fraction of probe muons which also meet this requirement gives an measurement of the efficiency.
In this way, the efficiency of reconstruction, trigger and isolation requirements are measured individually, and parameterized in terms of the geometry of the detector.
The method is repeated on simulated events, where results typically agree with data to within 3\%, and factors are applied to the simulated events to correct for any such discrepancies.
However, the muon trigger is not simulated; instead the trigger efficiency measured in data, with an average efficiency of approximately 88\%, is applied on an event-by-event basis to the simulated events.

The binning used for the data is selected based on a combination of detector resolution and data statistics considerations.
Detector effects on the resulting \ptz\ distribution are assessed by comparing the \ptz\ defined in terms of particles entering the detector to the \ptz\ reconstructed by the detector.
First, we define the dimuon system at the particle level in a way that can be implemented in any simulation.
We consider all particles with lifetimes $>10$~ps to have reached the detector.
From this list of particles, all muons with $\pt>15$~GeV and $|\eta|<1.7$\ are selected (regardless of their source in the generator event record), matching the detector acceptance.
By construction, the muons are considered after QED FSR, as would be measured in the tracking detector. 
Then, all possible opposite charge muon pairs are formed, and any which lie within the required mass range of 65--115~GeV are kept.
In the rare cases ($<0.5$\%) of events with more than one selected pair, the pair with mass closest to the $Z$~boson mass is used.
The requirement of non-colinearity of muon tracks applied at detector level is found to reject less than 0.2\% of particle level candidates, so is not applied at the particle level.
Similarly, the muon isolation and vertex requirements are treated as a detector level selection which is corrected for, and are not implemented at the particle level. 

\begin{figure}[!htb]\center
\includegraphics[width=80mm]{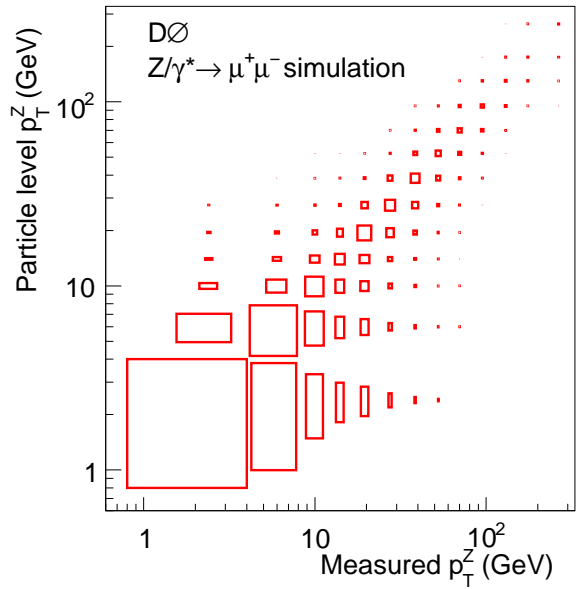}
\caption{\label{fig:ptz-matrix}The resolution matrix for \ptz\ from simulated events, used in correcting the measured data to particle level. The box area represents the number of events in a given particle level \ptz\ bin that are in a given bin of measured (detector level) \ptz. }
\end{figure}

We next correct the measured \ptz\ in data to the particle level, using the \alpgen+\pythia\ \zmm$+X$\ sample.
There are three possible scenarios for any given event, resulting in a three step process.
In the first scenario, a dimuon pair may pass all detector level selections, but fail one or more particle level selections. 
This class of events is dominated by migrations into the selected mass or muon $|\eta|$ regions due to detector resolution. 
In the simulated events, this class makes up approximately 2\% of the final sample with negligible dependence on \ptz, and this predicted contribution is subtracted from the measured data.
In the second scenario, the dimuon pair passes both the particle and detector level selections.
For these simulated events, the particle level \ptz\ is plotted against the detector level \ptz\ to assess the impact of detector resolution (see Fig.\ref{fig:ptz-matrix}).
The data distribution after the subtraction described in the first step is then corrected using a regularized inversion of this resolution matrix~\cite{guru}.
The regularization imposes the condition that second derivatives be small, which produces a smooth distribution; this smoothing is accounted for when deriving the uncertainties. 
In the third scenario, the particle level dimuon pair may pass selections, but the detector level pair fail selections. 
This effect is also assessed using simulated events and is dominated by inefficiency in the trigger or reconstruction and by gaps in detector coverage within the muon acceptance.
The data distribution resulting from the second step is corrected for these inefficiencies, parametrized as a function of (particle level) \ptz, giving the \z$+X$\ differential cross section.

Ensemble testing is used to determine the uncertainty on the differential cross section and any biases in the correction process.
The \pythia\ \zmm$+X$\ sample is used to build pseudo-datasets, after first being adjusted to describe the data by applying the ratio of the fully corrected data \ptz\ to the particle level \ptz\ in this \pythia\ sample.
Three hundred pseudo-datasets are drawn, with events chosen at random for each pseudo-dataset with a probability set so the average pseudo-dataset size matches the measured dataset.
Each pseudo-dataset is then treated exactly as data, and the detector level distribution corrected using the three step process described above.
The resulting corrected distribution is compared to the true particle level \ptz\ in that pseudo-dataset, and the fractional difference ($r_i$) is calculated for each \ptz\ bin. 
This process is repeated for all 300 pseudo-datasets. 
In a given \ptz\ bin, the 300 $r_i$\ form a Gaussian distribution and any shift of the mean of this distribution away from zero indicates a bias in the correction process.
The data are corrected for such biases, which are all at or below the 1\% level, and the uncertainty on the Gaussian mean is assigned as a systematic uncertainty.
The RMS of the $r_i$ distribution results from comparing the corrected distribution for each pseudo-dataset, which is smoothed by the regularization in the matrix inversion step, to the true particle level for that pseudo-dataset, which contains statistical fluctuations.
This RMS is therefore assigned to the data points as the statistical uncertainty. It is comparable to, but always larger than, 1/$\sqrt{N_{\text{data}}}$, where $N_{\text{data}}$\ is the number of detector level data events in a given bin.

Finally, further systematic uncertainties are assessed.
Several sources are considered, including the muon momentum resolution and the absolute momentum scale, and various detector efficiencies for muons.
Each source is shifted individually up and down by one standard deviation in the \alpgen+\pythia\ sample used to derive the correction factors, and left unchanged in the pseudo-datasets. 
The full ensemble testing process is repeated.
Statistical uncertainties are unaffected by these tests, but any change in the correction bias is assigned as a systematic uncertainty.
All systematic uncertainties are combined in quadrature, with the muon 1/\pt\ resolution being the largest source at low \ptz.
At high \ptz, detector efficiency effects are the largest systematic source, but the statistical uncertainties dominate.

Finally the differential cross section is normalized to the total dimuon production cross section (with the same muon $|\eta|$, \pt\ and dimuon mass requirements) measured in the data, determined by integrating over all \ptz. 
The dominant systematic uncertainties on the total cross section arise from the absolute determinations of the luminosity and muon trigger efficiency. 
Neither of these has a dependence on \ptz, and therefore they do not contribute to the uncertainty on the normalized distribution.

\begin{figure}[!htb]\center
\includegraphics[width=80mm]{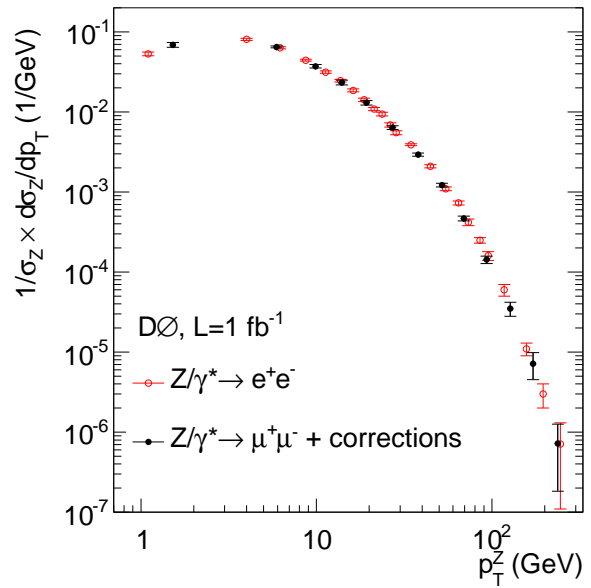}
\caption{\label{fig:comparison}Measurements of the normalized differential cross section in bins of \ptz\ for the dielectron~\cite{d0_zpt} and dimuon channels. Both results are shown with combined statistical and systematic uncertainties.}
\end{figure}

Table~\ref{tab:xsec} lists the normalized differential cross section, $(1/\sigma) \times (\text{d}\sigma/\text{d}p_T^Z)$, together with the statistical and systematic uncertainties.
We note that, due to the smoothing introduced by the regularization condition imposed during the second step of the data corrections, statistical fluctuations in the measured cross section in each \ptz\ bin have been suppressed; however the statistical uncertainties still accurately reflect the possible spread in each bin which could be caused by such fluctuations.
As a result, care must be taken when using the data in any fits as this suppression of fluctuations may lead to an artificially low $\chi^2$\ for any fit which describes the central values of the data well.
Table~\ref{tab:xsec} also lists four multiplicative correction factors for each bin, which can be applied to compare this result to previous measurements: the factor labeled $p_T^\mu$ corrects for the effect of the muon \pt$>15$~GeV requirement; the factor labeled FSR corrects for QED FSR;  the factor labeled $\cal A$\ then corrects from the measured lepton acceptance to full 4$\pi$ acceptance; and finally the factor labeled $\cal M$\ corrects from the measured mass window to the larger mass window used in the D0 electron channel measurement~\cite{d0_zpt} (40--200~GeV).
Applying only the $p_T^\mu$\ factor results in the same dimuon definition as previous \z+jets measurements~\cite{my_zjet, zjet_angles}; unlike \ptz, the variables studied in these previous measurements had minimal dependence on the muon \pt\ requirement, so a correction was applied by default.
All factors are derived using \resbos\ interfaced to \photos~\cite{photos}, as described in the following text, and we provide only the central values without assessing possible systematic uncertainties. 
However, deriving the same factors from the different theoretical calculations described in the following text indicates that model dependence limits the accuracy of these factors, particularly $\cal A$\ for $\ptz > 20$~GeV, to the level of a few percent.
Applying all factors to the data allows a comparison to the D0 electron channel measurement, as shown in  Fig. \ref{fig:comparison}. 
Within the limitations of this comparison, the agreement is reasonable.
For direct comparisons with theory, these correction factors are not applied to the dimuon data.

To compare to the data, predictions for the \ptz\ distribution are obtained from several theoretical calculations.
Predictions from pQCD are obtained with \mcfm, by evaluating both the differential distribution and total cross section at either leading order (LO) or next-to-leading order (NLO):
\begin{equation*}
f(\pt)\equiv\left.\frac{1}{\sigma_{Z/\gamma*}}\right|_{\text{(N)LO}}  \times \left.\frac{\text{d} \sigma_{Z/\gamma*}}{\text{d}\pt}\right|_{\text{(N)LO}}
\end{equation*}
where the first term is of order ${\cal O}(\alpha_s^0)$ at LO and ${\cal O}(\alpha_s^1)$ at NLO, while the second term is ${\cal O}(\alpha_s^1)$ at LO and ${\cal O}(\alpha_s^2)$ at NLO.
This approach differs from the treatment of the  pQCD calculation in the D0 electron channel measurement. There, both the total cross section and differential distribution were calculated to the same power of the strong coupling constant,  ${\cal O}(\alpha_S^2)$, yielding a NNLO  total \z\ cross section (and was labeled ``NNLO''), but a NLO differential distribution.
The prescription used here, calculating both the total and differential cross section to the same number of contributing terms in the perturbative expansion, results in a reduced scale dependence and improved convergence of the perturbative series~\cite{seymour}. 
The total cross section is evaluated using the inclusive $p\bar{p}\rightarrow\z+X$\ process at LO and NLO, and the differential distribution evaluated using the $p\bar{p}\rightarrow\Z+$~jet$+X$\ process again at LO and NLO, with no limit on the jet rapidity but requiring jet $\pt>2.5$~GeV to remove the divergence as $\pt\rightarrow0$. 
The same requirements are placed on the muons as for the data analysis, and the differential \z\ \pt\ distribution close to the jet \pt\ cutoff is excluded.
The \mstw2008 LO and NLO PDFs~\cite{mstw} are used throughout in calculating the LO and NLO processes respectively.
In all cases, renormalization and factorization scales are set to the sum in quadrature of the mass and \pt\ of the \Z\ in each event, and the dependence on this choice is assessed by varying both scales simultaneously up and down by a factor of 2, both for the differential distribution and the inclusive \z\ cross section used in normalization.
PDF uncertainties are assessed using the \mstw2008 68\%\ error sets, again taking into account the effect on the differential distribution and the inclusive \z\ cross section used in normalization. 
These are found to be approximately a factor of two smaller than the scale uncertainties at NLO, and negligible compared to the scale uncertainties at LO.
The prediction from \mcfm\ must then be corrected for the effects of QED FSR from the muons.
These corrections are derived from the \resbos+\photos\ sample described below, by comparing the distribution obtained by constructing the dimuon state using the muons before and after QED FSR, and are the inverse of the per-bin FSR corrections listed in Table~\ref{tab:xsec}.
We note that applying the pQCD prescription from the D0 electron channel measurement here would change the LO prediction to LO/NLO, and the NLO prediction to NLO/NNLO.
Differing only in the total cross section used to normalize, this change would lower the current predictions by  $28\%$ and $4\%$ respectively across all \z\ \pt, and increase the scale uncertainties by $\sim10\%$ (a $\le1\%$ absolute increase in the uncertainty).

A prediction for $(1/\sigma) \times (\text{d}\sigma/\text{d}p_T^Z)$ is also obtained from \resbos, using the \cteq6.6 parton distribution functions (PDF)~\cite{cteq66}.
At low \ptz, \resbos\ performs a next-to-next-to-leading logarithm (NNLL) resummation calculation, using the BLNY parametrization~\cite{blny}, with the default settings taken here. 
At higher \ptz\  ($\ptz\ge20$~GeV), \resbos\ transitions to an ${\cal O}(\alpha_s)$ pQCD calculation. 
Across the entire \ptz\ range, k-factors are applied in bins of \z\ \pt\ and rapidity to normalize to an ${\cal O}(\alpha_s^2)$ pQCD calculation~\cite{resbos_pqcd}.
The renormalization and factorization scales for the pQCD calculation are set to the mass of the \z\ in each event.
\resbos\ is interfaced to \photos\ for the simulation of QED FSR. 

Predictions for $(1/\sigma) \times (\text{d}\sigma/\text{d}p_T^Z)$ are also obtained from four event generators. 
Previous measurements~\cite{my_zjet, henrik_zjet, zjet_angles} indicate that the best description of boson+jets final states is currently provided by LO $2\rightarrow N$\ matrix element calculations with matched parton showers, as implemented in \sherpa\ and \alpgen, so we focus on these.
We use the same PDF set for all event generators: CTEQ6L1~\cite{cteq6l1}. 
First, a sample of events is generated with \sherpa, which uses the \comix\ matrix element generator~\cite{Gleisberg:2008fv} interfaced to a Catani-Seymour subtraction based parton shower~\cite{Schumann:2007mg}. 
Here, up to three partons are included in the matrix element calculation, and the threshold for matching to the parton shower is set to the default value of 30~GeV.
In \sherpa, the scales are determined dynamically during the matching process~\cite{Hoeche:2009rj}.
A sample of events is then generated with \alpgen, again with up to three partons in the matrix element calculation. 
The factorization scale is set to the sum in quadrature of the mass and \pt\ of the \Z\ in each event, and the renormalization scale set according to the CKKW prescription~\cite{ckkw}.
Parton jets from the matrix element calculation are required to have \pt$> 13$~GeV, $\Delta R ($jet, jet$) > 0.4$, and are limited to $|\eta|<2.5$.
These events are hadronized in three ways: first, using  \herwig\ (using an angular ordered parton shower) with \jimmy~\cite{jimmy} for multiple parton interactions, then using \pythia\ with underlying event tune D6~\cite{tune_D6}  (using the virtuality-ordered shower), and finally using \pythia\ with tune Perugia 6~\cite{perugia} (using the \pt-ordered shower~\cite{pt_ordered}).
This results in three different \alpgen\ predictions. 
In each case the default matching procedure is applied after hadronization,  requiring a $\Delta R ($jet, jet$) < 0.4$ match between parton jets and particle jets with $\pt>18$~GeV.
To determine the impact of the matching to the \alpgen\ matrix elements calculation, \herwig\ and \pythia\ are also tested directly in the same configurations described above:
 \herwig\ with \jimmy\ for multiple parton interactions, \pythia\ with tune D6, and \pythia\ with the \perugia\ tune.
In these configurations, all final state partons are generated by the parton shower. 
The renormalization and factorization scales for the hard scatter are set to the mass of the \z\ in each event, and are determined dynamically for the initial and final state showers.
For the \resbos, \sherpa, \alpgen, \pythia\ and \herwig\ calculations, the particle level quantities are extracted as defined earlier and each differential cross section prediction is normalized to the prediction of the dimuon cross section (with the same muon \pt, $|\eta|$\  and dimuon mass requirements) from that same model.

\begin{figure*}[!htb]\center
\includegraphics[width=160mm]{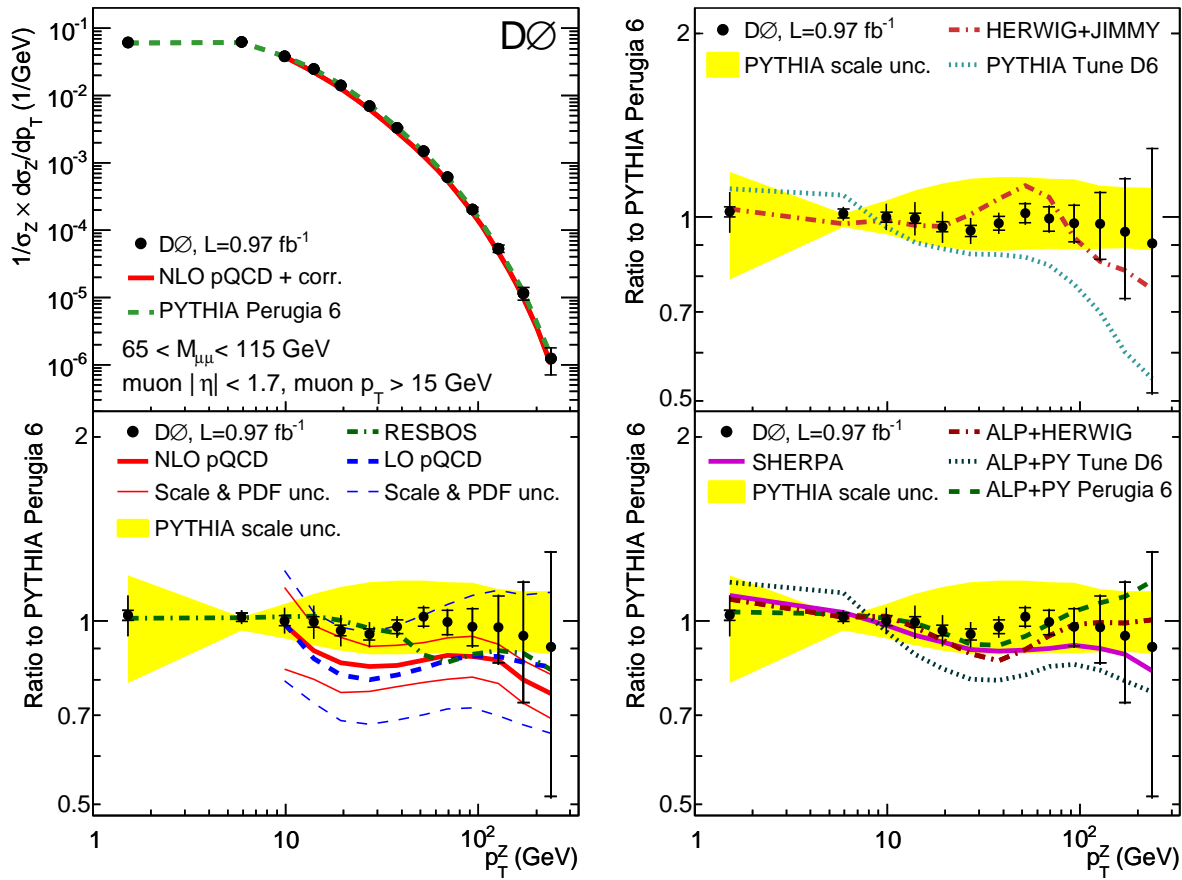}
\caption{\label{fig:result}The normalized differential cross section in bins of \ptz\ for $\Z(\rightarrow\mu\mu)+X$\ events. The data are shown with statistical uncertainties (horizontal bar) and combined statistical and systematic uncertainties (full bar). The distribution is shown in (a) and compared to fixed order and resummation calculations in (b), parton shower generators in (c), and the matrix element + parton shower generators in (d). All ratios in (b), (c), and (d) are shown relative to \pythia\ Perugia 6.}
\end{figure*}

The normalized differential cross section is presented in Fig.~\ref{fig:result}.
The data points are placed at the bin average, defined as the point where the differential cross section within the bin, taken from \pythia\ reweighted to match the shape in data, is equal to the measured value in the bin~\cite{bins}.
For clarity, only the predictions of NLO pQCD and \pythia\ Perugia 6 are shown with the data in  Fig.~\ref{fig:result}(a).
In the other parts of Fig.~\ref{fig:result}, ratios are shown.
To avoid repeating the data uncertainties and statistical fluctuations several times, we do not use data as the denominator in these ratios.
Instead, we choose \pythia\ Perugia 6, as this provides the best overall description of the data,  simplifying the determination of trends in other theoretical predictions relative to the data.
As an example of the scale uncertainty in an event generator, two further \pythia\ Perugia 6 samples are generated, with the scale for initial state QCD radiation varied up and down by a factor of 2.  
The effect of this change is shown as a shaded band around unity, and  shifts the distribution in opposite directions at low and high \ptz, with the transition point at approximately 6~GeV.
Further, even though \pythia\ is based on LO matrix elements, the scale uncertainty obtained is comparable to that on the NLO pQCD calculation, suggesting a cancellation of some of this scale variation in \pythia\ through a detailed balance between the matrix element for \z\ production and the Sudakov form factors from the parton shower. 
However, this small uncertainty is somewhat deceptive, as \pythia\ does not include a full NLO calculation.

\begin{table*}[h]
\caption{\label{tab:xsec}The measured normalized cross section in bins of dimuon \pt\ (\ptz) for \zmm$+X$\ events.
For each bin, we present the bin boundaries, the bin average ($\langle\ptz\rangle$, defined in the text), the normalized differential cross section, and the uncertainties.
Uncertainties are split into statistical, systematic with no correlations between bins (``uncorr. unc.''), and sources of systematic uncertainty that are correlated across all bins.
The sources of correlated uncertainty are: 1) muon $1/\pt$ resolution, 2) muon detection efficiency, 3) efficiency of all other selections. Factors to correct the result for the muon \pt\ requirement ($p_T^\mu$), QED FSR, muon acceptance ($\cal A$), and to the larger mass window of 40-200~GeV used in the D0 electron channel analysis ($\cal M$) are also provided (defined in the text).
}
{\begin{tabular}{r@{.}l@{--}r@{.}lr@{.}lr@{.}lr@{.}lr@{.}lr@{.}lr@{.}lr@{.}lr@{.}lr@{.}lr@{.}lr@{.}l}
\hline
\hline
\multicolumn{3}{l} \mbox{$p_T^Z$$\phantom{0}$} & & \multicolumn{2}{r} \mbox{$\langle\ptz\rangle$~~} & \multicolumn{2}{l} \mbox{$100\times\frac{1}{\sigma}\frac{d\sigma}{dp_T^Z}$ } & \multicolumn{2}{l} \mbox{Stat.~~} & \multicolumn{2}{l} \mbox{Uncorr.~~} & \multicolumn{2}{l} \mbox{Source 1} & \multicolumn{2}{l} \mbox{Source 2} & \multicolumn{2}{l} \mbox{Source 3} & \multicolumn{2}{l} \mbox{$p_T^\mu$~~~} & \multicolumn{2}{l} \mbox{FSR} & \multicolumn{2}{l} {~~$\cal A$} & \multicolumn{2}{l} {~~$\cal M$} \\

\multicolumn{3}{l} \mbox{(GeV)} & & \multicolumn{2}{l} \mbox{~~~(GeV)~} &  \multicolumn{2}{l} \mbox{(1/GeV)~~~~} & \multicolumn{2}{l} \mbox{unc. (\%)} &  \multicolumn{2}{l} \mbox{unc. (\%)} & \multicolumn{2}{l} \mbox{(\%)~~~} &  \multicolumn{2}{l} \mbox{(\%)~~~} &  \multicolumn{2}{l} \mbox{(\%)~~~}  \\
\hline
0&0~ & 4&0 & ~~1&5 & 6&13      & $\pm$1&2  & ~~$\pm$0&1 & ~~$\mp$7&2 & ~~$\pm$1&4 &~~$\pm$1&1 & 0&993 & 1&06 & 1&05 & ~1&02 \\
4&0~ & 8&0 & ~~5&9 & 6&27      & $\pm$1&0  & ~~$\pm$0&1 & ~~$\mp$1&5 & ~~$\pm$0&6 &~~$\pm$1&2 & 0&993 & 1&01 & 1&03 & ~1&01 \\
8&0~ & 12& & ~~9&9 & 3&84      & $\pm$1&2  & ~~$\pm$0&1 & ~~$\pm$4&3 & ~~$\pm$1&1 &~~$\pm$1&3 & 0&995 & 0&97 & 1&01 & ~1&00 \\
12&  & 16& & ~~14& & 2&50      & $\pm$1&3  & ~~$\pm$0&1 & ~~$\pm$5&9 & ~~$\pm$0&7 &~~$\pm$1&1 & 0&998 & 0&95 & 0&99 & ~0&99 \\
16&  & 23& & ~~19& & 1&43      & $\pm$1&3  & ~~$\pm$0&1 & ~~$\pm$5&4 & ~~$\pm$0&6 &~~$\pm$1&1 & 1&003 & 0&98 & 0&97 & ~0&98 \\
23&  & 32& & ~~27& & 0&704     & $\pm$1&5  & ~~$\pm$0&1 & ~~$\pm$4&2 & ~~$\pm$0&8 &~~$\pm$1&1 & 1&012 & 0&99 & 0&94 & ~0&97 \\
32&  & 45& & ~~38& & 0&332     & $\pm$1&9  & ~~$\pm$0&1 & ~~$\pm$2&7 & ~~$\pm$1&1 &~~$\pm$1&1 & 1&026 & 1&00 & 0&90 & ~0&96 \\
45&  & 60& & ~~52& & 0&150     & $\pm$2&9  & ~~$\pm$0&2 & ~~$\pm$1&4 & ~~$\pm$1&6 &~~$\pm$1&1 & 1&042 & 1&01 & 0&81 & ~0&95 \\
60&  & 80& & ~~69& & 0&0611    & $\pm$3&8  & ~~$\pm$0&3 & ~~$\pm$0&2 & ~~$\pm$2&3 &~~$\pm$1&1 & 1&059 & 1&02 & 0&75 & ~0&95 \\
80&  &~110& &~~93& & 0&0203    & $\pm$6&0  & ~~$\pm$0&6 & ~~$\mp$0&6 & ~~$\pm$2&7 &~~$\pm$1&6 & 1&081 & 1&03 & 0&67 & ~0&94 \\
110& &~150& &~~130& & 0&00530  & $\pm$11&0 & ~~$\pm$1&1 & ~~$\mp$1&2 & ~~$\pm$3&3 &~~$\pm$2&2 & 1&121 & 1&03 & 0&60 & ~0&94 \\
150& &~200& &~~170& & 0&00116  & $\pm$20&7 & ~~$\pm$1&4 & ~~$\mp$1&7 & ~~$\pm$3&7 &~~$\pm$2&4 & 1&143 & 1&04 & 0&55 & ~0&95 \\
200& &~330& &~~240& & 0&000123 & $\pm$42&0 & ~~$\pm$2&7 & ~~$\mp$2&1 & ~~$\pm$4&1 &~~$\pm$2&6 & 1&131 & 1&05 & 0&52 & ~0&95 \\
\hline
\end{tabular}}
\end{table*}

Comparisons to the data indicate two regions: $\ptz<30$~GeV, where the resummation calculation provides a good description of the data, and $\ptz>30$~GeV, where the fixed order calculation provides the best description. 
In this higher \ptz\ region, the NLO pQCD calculation is a significant improvement in uncertainty over LO, however an overall normalization difference relative to the data is observed.
For $30<\ptz<100$~GeV, this difference is between 1--2 standard deviations of the combined data and theory uncertainties, with the theory scale uncertainty dominating and the choice of a lower scale bringing the pQCD calculation into better agreement with the data.
For $\ptz>100$~GeV, the data statistical uncertainty dominates, and the theory remains below the data but is consistent within this uncertainty.
This disagreement with pQCD predictions is in the same direction as observed in the D0 electron channel measurement, but significantly smaller. 
However, the detector acceptance for the electron channel was larger than for the muon channel, and attempting to extrapolate between these acceptances revealed a dependence on the theoretical models used, complicating direct comparisons of the two results. 
These two measurements in fact provide different information on the \ptz\ distribution in different \z\ rapidity ranges, and future measurements which further probe the correlations between the \z\ \pt\ and rapidity are clearly of interest and may further illuminate the disagreements seen when comparing pQCD to data.

Of the event generators, \pythia\ \perugia\ provides the best description of the data over the full \ptz\ range, and we note that the D0 electron channel measurement was used as an input in deriving this tune.
All other event generators agree within the combined theory and data uncertainties, except the \pythia\ D6 and \alpgen+\pythia\ D6 predictions.
Interfacing \pythia\ and \herwig\ with \alpgen\ clearly affects the region dominated by the \alpgen\ matrix elements, though the agreement with data is equally good.
While the \sherpa\ prediction agrees with the data within uncertainties, it generally follows the shape of the \pythia\ \perugia\ prediction with a higher scale choice suggesting that, as for the pQCD calculation, a lower scale choice in \sherpa\ may yield an even better description of the data.

In summary, we have presented a new measurement of the normalized \Z($\rightarrow\mu\mu$)+$X$\  cross section, differential in the dimuon \pt.
This is the first such measurement at the level of particles entering the detector, allowing unbiased tests of theoretical predictions.
The measurement was made using a sample corresponding to  $0.97$~fb$^{-1}$\ of integrated luminosity recorded by the D0 experiment in \pp\ collisions at \roots. 
The current best predictions for vector boson production at hadron colliders were tested, and these predictions have varied success in describing the data. 
In particular, the disagreement with pQCD seen in the electron channel measurement at high \z\ \pt\ is smaller within the kinematic acceptance of this measurement, and the use of a lower scale within the calculation further reduces the disagreement. 
An accurate description of both the low and high \ptz\ regions is also essential in predicting the production rates and kinematics of jets in association with the \z, and this result is an important input for the tuning of theoretical predictions.
Improving the modeling of this process will lead to increased sensitivity of searches for rare and new physics.

%
We thank Zhao Li, Mike Seymour, Frank Siegert, Peter Skands and Chien-Peng Yuan for very useful discussions on the theoretical calculations.
We thank the staffs at Fermilab and collaborating institutions,
and acknowledge support from the
DOE and NSF (USA);
CEA and CNRS/IN2P3 (France);
FASI, Rosatom and RFBR (Russia);
CNPq, FAPERJ, FAPESP and FUNDUNESP (Brazil);
DAE and DST (India);
Colciencias (Colombia);
CONACyT (Mexico);
KRF and KOSEF (Korea);
CONICET and UBACyT (Argentina);
FOM (The Netherlands);
STFC and the Royal Society (United Kingdom);
MSMT and GACR (Czech Republic);
CRC Program and NSERC (Canada);
BMBF and DFG (Germany);
SFI (Ireland);
The Swedish Research Council (Sweden);
and
CAS and CNSF (China).


\begin{thebibliography}{99}
\bibitem{eta} $\eta = -\ln [\tan(\theta/2)]$, where $\theta$\ is the polar angle, defined with respect to the proton beam direction.
\end{thebibliography}
\end{document}